\documentclass[12pt]{article}
\usepackage{graphicx}
\def\degr{\hbox{$^\circ$}}
\def\xmm{XMM-{\it Newton}}

\title{\bf CPD\,$-$41\degr7742: an unusual wind interaction\footnote{Based on data collected at the European Southern Observatory (La
Silla, Chile) and with \xmm, an ESA Science Mission with instruments and 
contributions directly funded by ESA Member States and by the USA (NASA). }}
\author{
 H. Sana$^1$\thanks{FNRS Research Fellow (Belgium)} , 
 E. Gosset$^1$\thanks{FNRS Research Associate (Belgium)} ,
 G. Rauw$^{1\ddagger}$,
 E. Antokhina$^2$,
 P. Royer$^3$,
 J. Manfroid$^1$\thanks{FNRS Research Director (Belgium)} 
 \\and J.-M. Vreux$^1$\\
\vspace{1cm}\\
\normalsize $^1$Institut d'Astrophysique et de G\'eophysique, Li\`ege University, All\'ee du 6 Ao\^ut 17, Bat. B5c, \\ \normalsize B-4000 Li\`ege, Belgium\\
\normalsize $^2$Sternberg Astronomical Institute, Moscow State University, Universitetskii pr., 13, \\ \normalsize 119899 Moscow, Russia\\
\normalsize $^3$Instituut voor Sterrenkunde, Katholieke Universiteit Leuven, Celestijnenlaan 200 B, \\ \normalsize B-3001 Leuven, Belgium\\ 
}
\date{\mbox{}}
\begin{document}
\maketitle
\pagestyle{empty}
%
%
\def\bull{\vrule height .9ex width .8ex depth -.1ex}
\makeatletter
\def\ps@plain{\let\@mkboth\gobbletwo
\def\@oddhead{}\def\@oddfoot{\hfil\tiny\bull\quad
``Massive Stars and High-Energy Emission in OB Associations''; JENAM
   2005, held in Li\`ege (Belgium)\quad\bull}%
\def\@evenhead{}\let\@evenfoot\@oddfoot}
\makeatother
%
%
\def\beginrefer{\section*{References}%
\begin{quotation}\mbox{}\par}
\def\refer#1\par{{\setlength{\parindent}{-\leftmargin}\indent#1\par}}
\def\endrefer{\end{quotation}}
%
%
{\noindent\small{\bf Abstract:} 

We summarize the results of a multiwavelength observing campaign on the massive eclipsing binary CPD\,$-$41\degr7742, another remarkable object in the young open cluster NGC 6231. Our campaign relies on high resolution echelle spectroscopy, narrow-band optical photometry, and \xmm\ X-ray observations. Combined with the spectroscopic analysis, the light curve analysis provides a direct measurement of the masses and sizes of the system components. However, the most outstanding results come from the \xmm\ observations. Our 180~ks campaign towards NGC 6231, and CPD\,$-$41\degr7742, provides an unprecedented phase coverage of such a close early-type binary. The EPIC-MOS light curves almost fully cover the 2.4 day period of the system and the brightness of the object is sufficient to yield a time resolution as tight as 1~ks.  The X-ray flux presents clear variations along the orbit, that we interpret as the signature of an unusual wind interaction. We indeed expect that, in this O+B system, the dominant primary wind crashes into the secondary surface, leading to a wind-photosphere interaction. As a strong support to our interpretation, we provide a geometrical model that associates an extra X-ray emission to the secondary inner surface. Though quite simple, the present model matches the main features of the X-ray light curve.
%
%

\section{Introduction}
CPD\,$-$41\degr7742\ is an SB2 eclipsing binary with a period $P\approx2.44$d. In Sana et al. (2005a), we analyze its optical light curve and, combining these new constraints with those obtained from the spectroscopy (Sana et al. 2003), we derive the absolute orbital and physical parameters of the system. In particular, we show that CPD\,$-$41\degr7742\  is most probably formed by an O9V and a B1V star. In the present contribution, we summarize the analysis of the \xmm\ EPIC-MOS X-ray light curve. The details of our observing campaign and of our analysis are to be found in Sana et al. (2005a).}

\begin{figure}
\centering
\includegraphics[width=7cm]{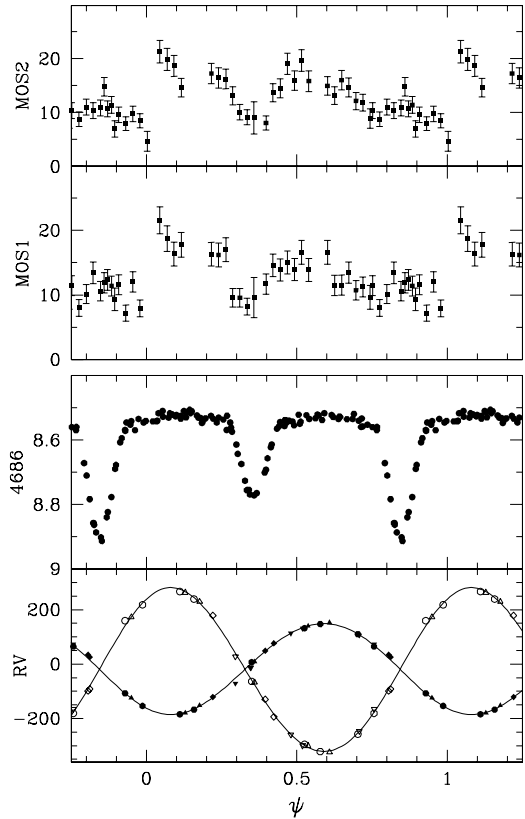}
\caption{
From bottom to top: RV curves (in km s$^{-1}$), optical light curve (in mag, 
at $\sim$~$\lambda$4686\AA) and \xmm\ EPIC-MOS1 and MOS2 (in $10^{-3}$ cts 
s$^{-1}$) X-ray light curves of CPD\,$-$41\degr7742. The bin size of the X-ray 
curves is 1~ks.
 }
\end{figure}

\section{An unprecedented X-ray light curve}

In Sept.\ 2001, the young open cluster NGC 6231 has been the target of a 180~ks monitoring campaign with \xmm\ (Sana et al. 2005b). These X-ray observations yielded an almost complete coverage of the CPD\,$-$41\degr7742\ orbital cycle. The brightness of the object was further sufficient to obtain a time resolution as tight as 1~ks. Briefly, \xmm\ revealed a clearly variable, though complex, X-ray light curve (Fig.~1). In particular, we observed an X-ray eclipse (at $\phi\approx0.35$) almost perfectly synchronized with the secondary eclipse in the optical light curve (Fig.~1). This suggests the presence of an additional localized X-ray component associated with the secondary inner surface. 

\section{An unusual wind interaction}
In CPD\,$-$41\degr7742, the winds from the two components are strongly off balance and no ram pressure equilibrium can be sustained on the binary axis. As a consequence, the overwhelming O-type primary wind most probably crashes into the B-type secondary surface. In this scenario, the extra X-ray emission component produced by the wind interaction is naturally associated with the secondary inner surface (Fig.~2).

\subsection{A wind-photosphere interaction model}
To test our interpretation of the CPD$-$41\degr7742 X-ray light curve, we built a simple phenomenological model based on the following assumptions:
\begin{enumerate}
\item[-] {\bf Ingredients:} 3-D geometry, spherically symmetric stars and winds, circular orbit, accelerated winds ($\beta = 1$ velocity law), neglect of the influence of the orbital motion, fully radiative shock (immediate cooling)

\item[-] {\bf Fixed parameters:} geometry (sizes, separation, inclination), constrained from the SB2E orbital solution; wind parameters ($\dot{M}$ and $v_{\infty}$), derived from the mass loss recipes of Vink et al. (2001)

\item[-] {\bf Free parameter:} amplitude of the emissivity (or, equivalently, efficiency of the cooling)
\end{enumerate}

To compare the {\it predictions} of our model with the observed X-ray light curve, we separate the signal into a constant emission threshold and a variable component with which we confront the modelled light curve. As seen from Fig.~3, our simple model reproduces to the first order the  modulations of the X-ray flux of CPD$-$41\degr7742. In particular, the width of the X-ray {\it eclipse} is well rendered. 

  \begin{figure}
  \centering
  \includegraphics[angle=90,height=5cm]{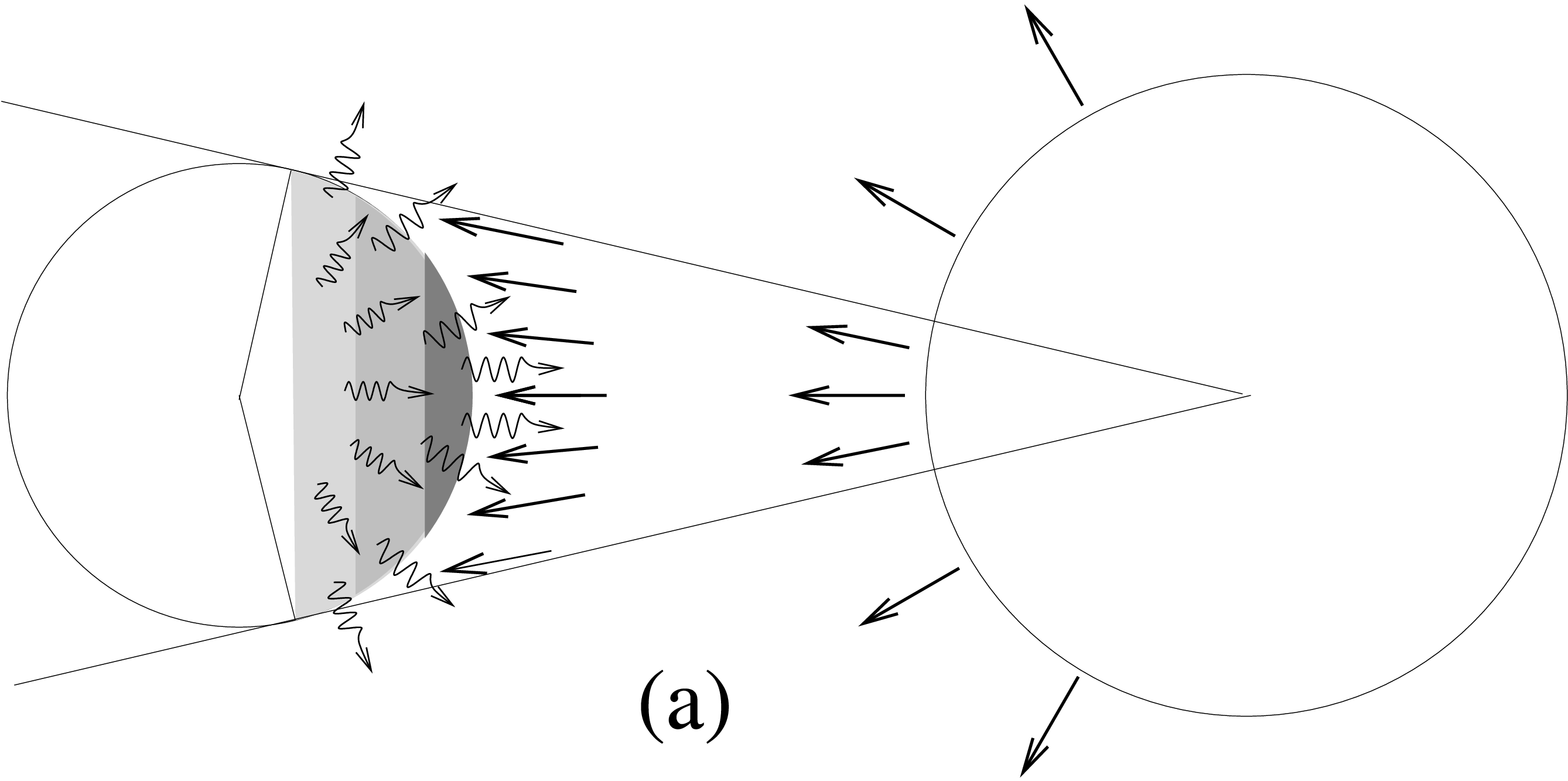} \hspace*{5mm}
  \includegraphics[height=4.5cm]{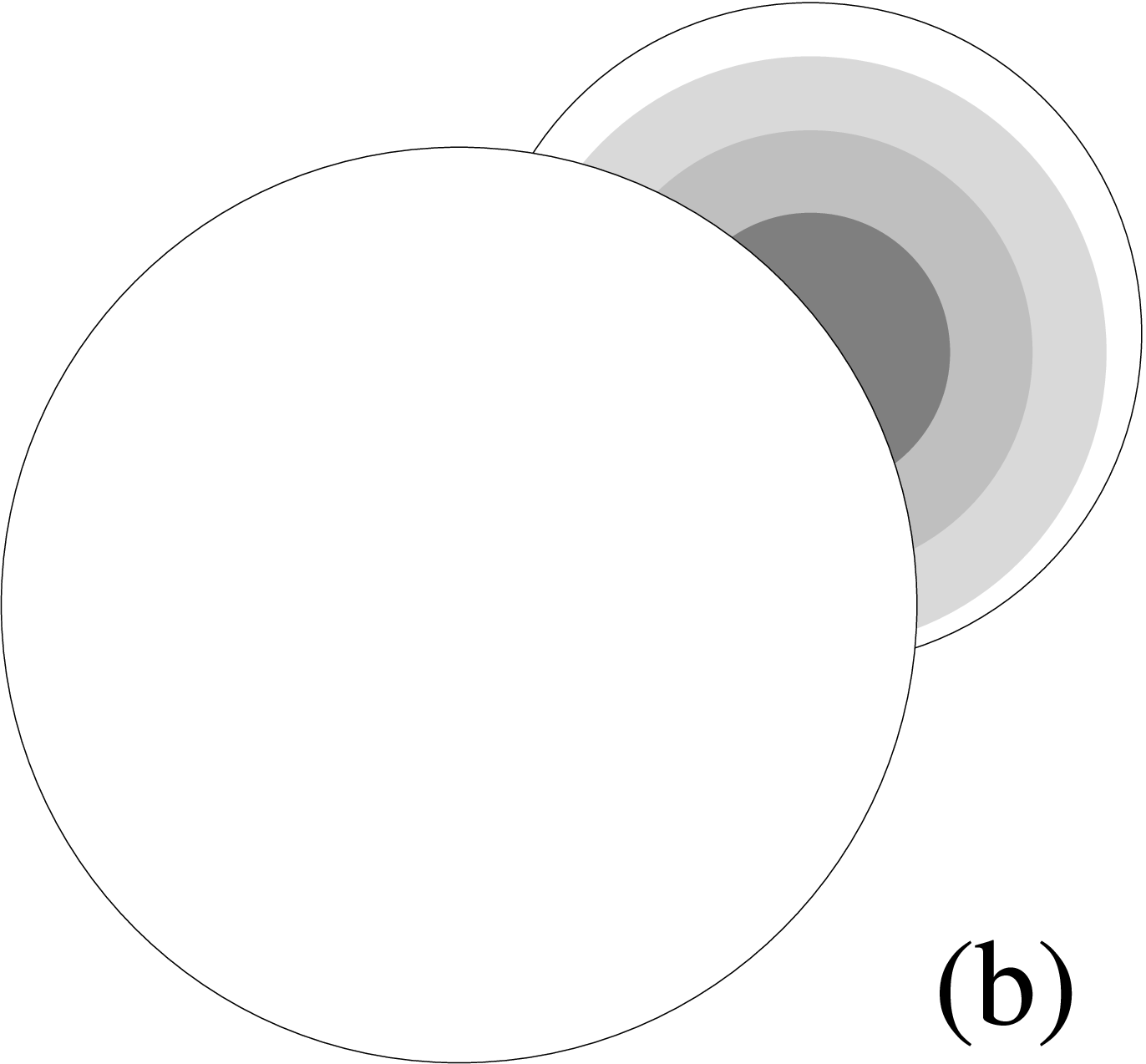}
  \caption{Schematic view of our geometric interaction model. {\bf (a)} A view from above the orbital plane: the secondary star intercepts a small fraction of the primary wind, of which part of the kinetic energy is turned into heat.  {\bf (b)} A similar view along the line of sight. The different degrees of shading of the secondary surface represent the different X-ray emissivities, these latter being larger closer to the system axis.}
  \end{figure}

\subsection{Final remark}

We can not rule out that radiative braking (Gayley et al. 1997) be strong enough to alter the wind-photosphere interaction, resulting in a wind-wind interaction structure instead. However, the latter should still be located close to the secondary surface. Briefly, even if braking occurs, we still expect similar modulations of the X-ray flux.

\begin{figure}
\centering
\includegraphics[width=7.9cm]{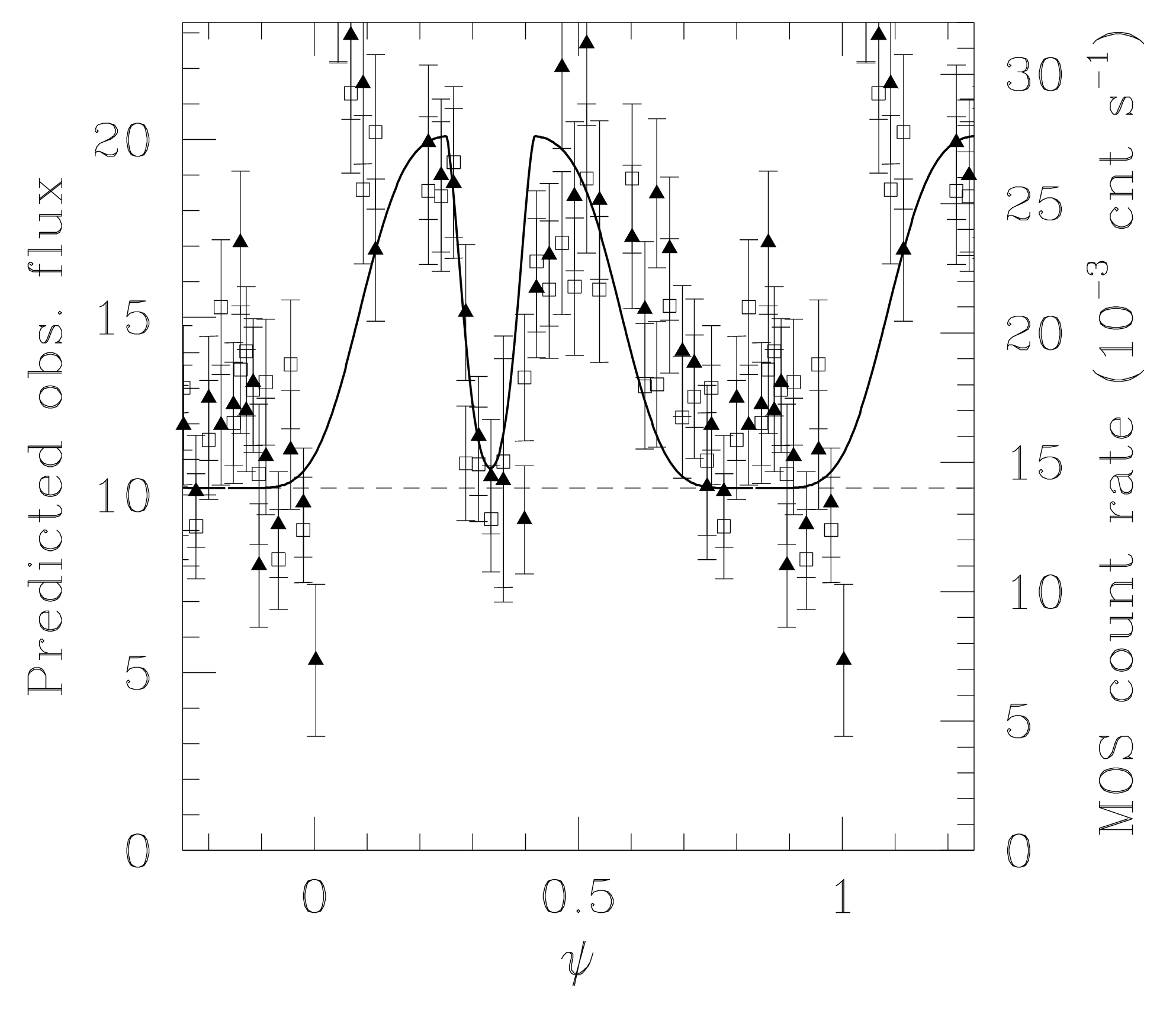}
\caption{
EPIC-MOS X-ray light curves. The thick line indicates the contribution 
of the wind interaction component computed from our geometrical model. 
The amplitude of the predicted variations has been arbitrarily scaled to
 the data.  }
\end{figure} 

\section{Conclusions} 
Our \xmm\ campaign almost fully covered the orbital cycle of the early-type binary CPD$-$41\degr7742 and yielded an unprecedented X-ray light curve of such a close system. The clear modulations of the X-ray flux are interpreted as the signature of a wind interaction, though of a particular kind: the overwhelming primary wind crashes into the secondary star, producing an extra X-ray component associated with the secondary inner surface. It is the first time that such a wind-photosphere interaction is unveiled from the observational point of view. Finally, the case of CPD$-$41\degr7742 outlines the need for phase-resolved observations of close early-type binary systems, that provide a phase-coverage as good as possible. It is only thanks to the unprecedented quality of the EPIC-MOS  X-ray light curve of CPD$-$41\degr7742 that we could unveil the unusual wind interaction happening in this system.

%
%
\section*{Acknowledgements}
The Li\`ege team acknowledges support from the PRODEX XMM and Integral Projects, as well as contracts P4/05 and P5/36 `P\^ole d'Attraction Interuniversitaire' (Belgium). EA acknowledges support from the Russian Foundation for Basic Research (project No 02-02-17524) and the Russian LSS (project
No 388.2003.2). 
%
%
 
\beginrefer
\refer Gayley K.G., Owocki S.P. \& Cranmer S.R. 1997, ApJ 475, 786

\refer Sana H., Hensberge H., Rauw G. \& Gosset E. 2003, A\&A 405, 1063

\refer Sana H., Antokhina E., Royer P., Manfroid J., Gosset E., Rauw G. \& Vreux J.-M. 2005a, A\&A 441, 213

\refer Sana H., Gosset E., Rauw G., Sung H. \& Vreux J.-M. 2005b, A\&A, in press

\refer Vink J.S., de Koter A. \& Lamers H.J.G.L.M. 2001, A\&A 369, 574

\endrefer           
\end{document}